\documentclass[letterpaper, a4paper]{amsart}

\usepackage{graphicx}
\usepackage{latexsym}
\usepackage[all]{xy}
\usepackage{color}

\usepackage[margin=1in]{geometry}

\numberwithin{equation}{section}
\begin{document}
\begin{title}[Quantum illumination with Non-Gaussian Three Photons States]{Quantum illumination with Non-Gaussian Three Photons States}
\end{title}
\date{\today}
\maketitle
\thispagestyle{empty}
\begin{center}
\address{
Fraunhofer Institute for High Frequency Physics and Radar Techniques FHR  \\
Department of Mathematics, University of Primorska}
\end{center}
\begin{center}
\author{Ricardo Gallego Torrom\'e\footnote{Email: rigato39@gmail.com}}
\end{center}

\begin{abstract}
It is shown that quantum illumination with three photons non-Gaussian states, where the signal is described by a two photons state and the idler is described by a one photon state, can outperform in sensitivity standard Gaussian quantum illumination in a high noisy background. In particular, there is a reduction in the probability due to an increase in the probability of error exponent by a factor $1/{N_S}$, where $N_S$ is the average number of photons per mode of the signal state.
\end{abstract}
\bigskip
{\small
{\bf Keywords:} Quantum Radar, Quantum Illumination, Quantum Enhancement, Non-Gaussian states.}
\section{Quantum illumination}
Quantum illumination is a quantum sensing protocol that exploits the benefits of inherited quantum correlations from entanglement to harvest an improvement in detection sensitivity \cite{Lloyd2008,Tan} or target detection accuracy \cite{Zhuang Shapiro 2022}. Quantum illumination is also a prominent example of quantum protocol with benefits against equivalent classical strategies, even if the initial entanglement is lost due to losses and background noise, making it likely to be implemented for technological applications without the concerns that plague other quantum protocols based upon quantum phenomena.

 The standard protocol of quantum illumination works according to the following scheme. From a source of entangled photon pairs and for each pair, one of the photons (signal photon) is sent to explore a region of spacetime, while the partner photon (idler) is keep alive in the laboratory system. When the return photon arrives to the receiver, a joint measurement received photon/idler states is performed.
In situations of high noise back-ground, quantum illumination with Gaussian quantum illumination of Tan et al. offers an advantage over coherent illumination of $6 dB$ in probability of error over the equivalent coherent state illumination \cite{Tan}. It was showed that, quantum illumination with two mode squeezed vacuum states, when idler and signal beams have the same intensities, provides an advantage of $6\,dB$ over coherent light illumination with the same characteristics of frequency and intensity beam \cite{Tan,Shapiro2019}. Furthermore, theoretical analysis have revealed that the TMVS quantum illumination provides the optimal strategy to minimize the probability of error \cite{de Palma Borregaard, Nair Gu 2020}. However, such studies limit their scope to two photon states.

Quantum illumination protocol is applicable at optical and microwave frequency domains \cite{Barzanjeh et al.} and several experimental demonstrations have confirmed the theoretical advantage over coherent light \cite{Zhang et al. 2015,England Balaji Sussman 20019, Barzanjeh et al.2019,Luong et al. 2019,Luong et al. 2020}.

\begin{figure}
\graphicspath{{C:\Users\rgall\Desktop\Ricardo 09-2019\Desktop\Ricardo backup 2018 05\Ricardo\trabajos mios, PhDs y tesinas/}}
\begin{center}
\includegraphics[scale=0.3]{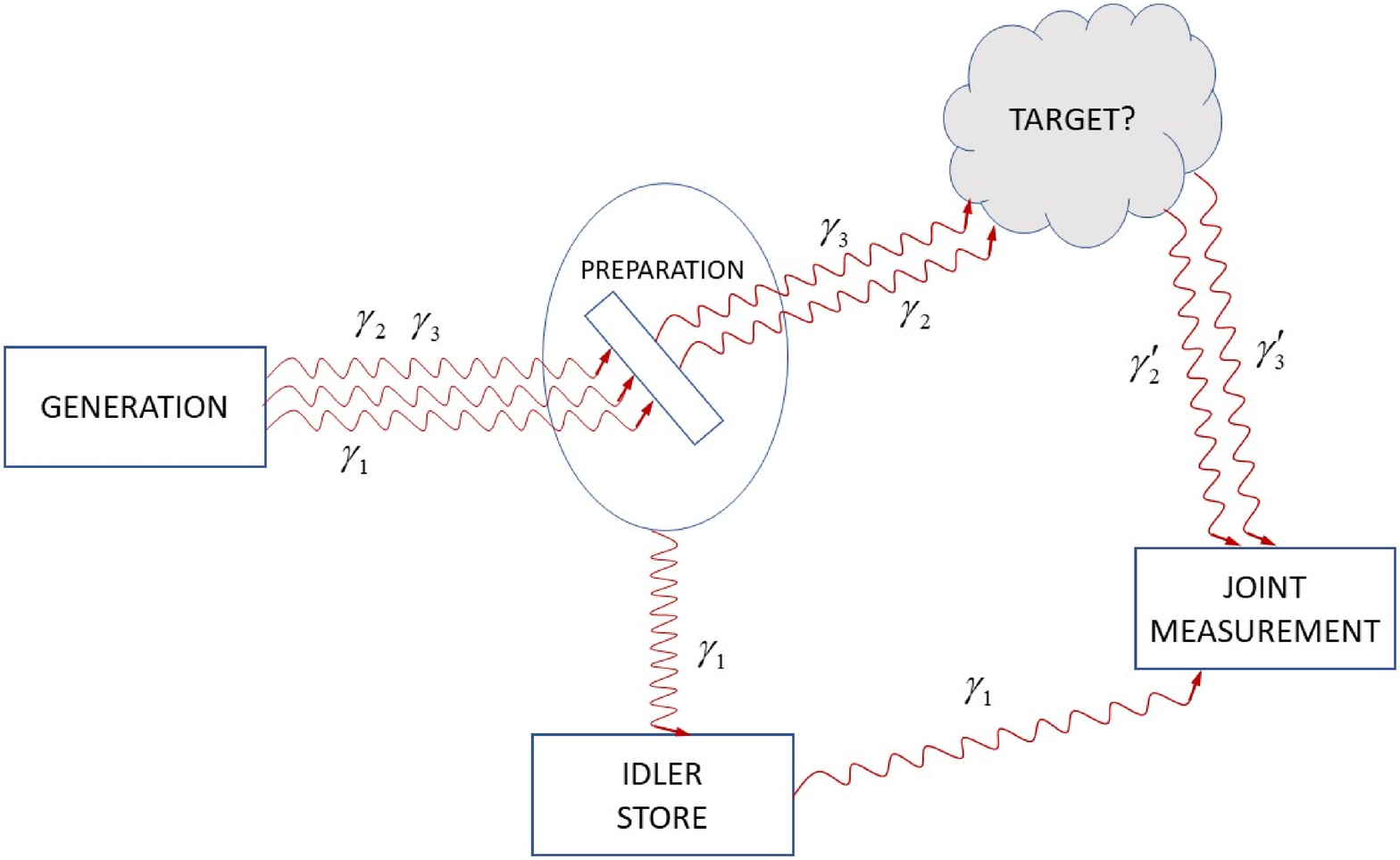}
\caption{{\small {\bf Quantum illumination with three entangled photons}. Three photons correlated in time and in frequency are generated. Two photons (signal photons) are sent to explore a region where a target could be located; one photon is either retained or recorded (idler photon). The two signal photons are simultaneously measured in a joint measurement with the idler photon.}}
\label{qi2p joint measurement}
\end{center}
\end{figure}

 The scheme of quantum illumination with multiple entangled photons was proposed in \cite{Ricardo 2021}, see Fig. \ref{qi2p joint measurement} for the case under study of three-photon states. From a source of three photon entangled states, two signal photons are sent to explore a region of spacetime where a possible target can be located, while the companion third photon is kept alive in the laboratory frame as idler photon. When the return pair of photons return, a joint measurement together with the idler photon is performed. Note that a detection is declared if two simultaneous received photons are detected and correlated with the idler photon.

In ref. \cite{Ricardo 2021}  quantum illumination with multiple entangled states was analyzed for the case of a direct generalization of Lloyd's quantum illumination \cite{Lloyd2008}. Such models, although of interest to conceptualize the main ideas of quantum illumination, are less interesting from a more practical point of view, because coherent illumination can offer a better performance \cite{ShapiroLloyd} and because, as in the case of atmospheric applications, the conditions assumed by the models, do not hold in generic realistic situations.

The aim of the present paper is to describe the sensitivity properties of certain generalization of quantum illumination that potentially overcome the applicability loopholes of the theory described in \cite{Ricardo 2021}. Recent advances in three-photon entangled states generation based on non-Gaussian states whose beams reach intensities comparable to the generation of two-photon states, has open the possibility to use non-gaussian entanglement \cite{C.W. Sandbo Chang et al. 2020}. Therefore, it is natural to ask wether the three photons states investigated in  \cite{C.W. Sandbo Chang et al. 2020} can be used for quantum illumination. In particular, we will consider in this paper if quantum illumination with such states non-Gaussian states have advantage with respect to standard Gaussian quantum illumination in terms of probability of errors.  Our result is that, similarly as the case studied in \cite{Ricardo 2021} showed an improvement with respect to Lloyd's protocol, the three-photons entangled states that phenomenologically describe \cite{C.W. Sandbo Chang et al. 2020}, the probability of error can be significantly smaller than for Gaussian quantum illumination \cite{Tan}. Indeed, in a reduced scope scenario of very high noise compare with the reflectivity of the target, the reduction in probability of error exponent is equal to the probability of error exponent of Tan et al. model but multiplied by factor $1/{N_S}$. This is translated a reduction in the probability of error, that for the conditions of low $N_S$, can decrease the probability of error dramatically.

\section{Non-Gaussian three photons states and Quantum Illumination}
The process of generation of three-photon  entangled  states that we consider here is a four wave mixing process in the form of a three-photon spontaneous parametric down conversion in the microwave domain, producing tri-squeezed states \cite{C.W. Sandbo Chang et al. 2020}. After averaging out highly oscillating terms, the effective interacting Hamiltonian for the process in question is of the form
\begin{align}
\widehat{H}^{3}_I =\,\hbar g_0\,\left(a_1 a_2 a_3\,+ a^\dag_1 a^\dag_2 a^\dag_3\right),
\label{Hamiltonian 3m}
\end{align}
where $g_0$ is a constant.
Specifically, we are interested in three-photon spontaneous parametric down conversion to multiple modes with frequencies $\omega_1$, $\omega_2$ and $\omega_3$. Such processes were experimentally realized by means of a {\it meandered design} SQUID, that allowed a reduction of the pump intensity required for the three-photon conversion up to three order magnitude than in previous experiments, a critical factor to enhance the required conversion.

The initial state associated to the pump flux is described by a thermal state. After the interaction is applied and up to third order in $gt$, the system is described by the state
\begin{align}
\widetilde{|\psi\rangle} =\,\left(1-\frac{1}{2}(gt)^2\right)|000\rangle\,-\imath gt\,|111\rangle .
\label{Approximate Non-gaussian state for 3m}
\end{align}
The state \eqref{Approximate Non-gaussian state for 3m} provides an effective theoretical description of the phenomenology observed \cite{C.W. Sandbo Chang et al. 2020}. Hence we adopt this state for our considerations.
Assuming that $\theta:=gt$ is small, a condition that is compatible with the above mentioned experiments, the parameter $\theta ^2$ is approximately equal to the average photon number per mode of the signal photons $n_s$ is given by the expression
\begin{align}
{n}_s :=\widetilde{\langle \psi|}\hat{a}^\dag_1\hat{a}_1\widetilde{|\psi\rangle}=\,\theta^2+\mathcal{O}(\theta^3).
\end{align}

We can  re-cast the state $\widetilde{|\psi\rangle}$ in a close form.
Up to third order in $\theta$, the state \eqref{Approximate Non-gaussian state for 3m} is equivalent to
\begin{align*}
|\psi\rangle =\,\cos (\theta)|000\rangle-\,\imath \sin (\theta)|111\rangle,
\end{align*}
that can be succinctly re-cast as
\begin{align}
|\psi \rangle =\,{\bf U}\,|000 \rangle  ,
\label{Estate for psi}
\end{align}
where the unitary transformation ${\bf U}:\,V\to V$ in the $2$-dimensional space $V$ spanned by $\{|000\rangle,|111\rangle\}$ is determined by the matrix
\begin{align*}
U:=\begin{pmatrix}
\cos (\theta) & -\imath \sin (\theta)\\
-\imath \sin(\theta) & \cos (\theta)
\end{pmatrix}.
\end{align*}
With this notation, the density matrix for the idler-signal system is of the form
\begin{align}
\rho_\psi(t) =\,{\bf U}|000\rangle\langle 000 |{\bf U}^\dag .
\label{density psi}
\end{align}
The idler state, when separated from the signal and if not altered, takes the form
\begin{align}
|I\rangle_1 = U |0\rangle_1.
\label{idler state}
\end{align}
that corresponds to the partial trace of $\rho_\psi(t)$ with respect to the signal photon subspaces.

Finally, the back-ground noise is modeled by thermal states. For the case that we are discussing, the back-ground state is the product of two thermal states, one that corresponds to the photon with frequency $\omega_1$ and the other to the photon with frequency $\omega_2$. The thermal state is of the form
\begin{align}
\rho_{th} =\,\sum^{+\infty}_{n=0} \frac{\bar{n}^n}{(\bar{n}+1)^{n+1}}|n\rangle \langle n |
\label{rhothermal}
\end{align}
where $\bar{n}$ is the is the average number of photons per mode as is a function of the frequency $\omega$.
 Since in the case that we are considering there are two simultaneous photons in the signal state, one needs to consider the product state $\tilde{\rho}_B=\,\rho_{th}(\omega_2) \otimes\,\rho_{th}(\omega_3)$. When $\bar{n}\gg \,1$, this state can be expressed as $\rho_{th}\simeq I_n/{\bar{n}}$. Thus the state describing the back-ground noise will be of the form,
 \begin{align}
 \rho_B \approx \,\frac{I}{\bar{n}_2}\otimes \frac{I}{\bar{n}_3},\quad I=\sum^{+\infty}_{k=0} |k\rangle\langle k | .
 \label{background state}
 \end{align}
 where $\bar{n}_2$ and $\bar{n}_3$ are the average numbers of photons per mode for the corresponding frequencies. Note the expression for $\rho_B$ in \eqref{background state} is a convenient approximation for $\rho_{th}\otimes\rho_{th}$. In order to have a consistent expression for the density it needs to be normalized to have trace equal to $1$. Not being directly normalized in the full infinitessimal Hilbert space of all possible modes, $I/{\bar{n}}$ is a multiple of a resolution of the unity. The substitution of $\tilde{\rho}_B$ by $\rho_B$ is very close to $1$ and being multiplicative of the form $(1+\alpha(\bar{n})/\bar{n})$, with $\alpha(\bar{n})/\bar{n}\to \, 0$ when $\bar{n}\to +\infty$, it will be approximated to $1$ in our considerations.

\section{Probability of error and comparison with two modes Gaussian quantum illumination}
The main task is to discriminate between two equally plausible situations, one where the target is absent, a possibility that is described by the state $\rho_0$), and the  second situation corresponding to the presence of a target, a possibility represented by the state $\rho_1$. One standard resolution of this problem makes use the quantum Chernoff bound. Let us assume a {\it positive operator valued measure} (POVM), that consists in this case of two operators $E_0,E_1$ such that $E_0+\,E_1=\,\mathbb{I}$ and $E_i \geq 0$. Given two possible quantum states with corresponding a priori assigned probabilities $\pi_0,\pi_1$, the  probability of error is given by the expression
 \begin{align}
 P_{err} = \,\pi_0\,Tr [E_1 \,\hat{\rho}_0]+\,\pi_1 \,Tr[E_0\,\hat{\rho}_1].
  \end{align}
  The basic problem to solve is to understand how the error probability behaves for $M$ experiments where the state can be either in state $\hat{\rho}_0$ (hypothesis $H_0$) or in the sate $\hat{\rho}_1$ (hypothesis $H_1$). The resolution in the asymptotic case of $M$ determines the quantum Chernoff's theorem, that for the case under consideration implies that the probability of error of making a binary decision after $M$ large independent measurements is determined by Chernoff's type bound,
 \begin{align}
P_{err}(M)\leq\, \exp(-M\,C_q),
 \end{align}
 where the exponent is given by the expression
 \begin{align}
 C_q =\,\lim_{M\to \infty}\,-\frac{\log (P_{err}(M))}{M}.
 \end{align}
 In the limit $M\to +\infty$, the bound is saturated and one has an equality.

It is useful to reformulate the above Quantum Chernoff's type bound as follows. The probability of error $P_{err}(M)$ is bounded as
 \begin{align*}
 P_{err}(M)\leq \,\frac{1}{2}\,\exp\left(-\alpha\,M\right),
 \end{align*}
where
\begin{align*}
\alpha=\,-\log \left(\inf_{s\in[0,1]}\,Tr\left(\rho^s_0\rho^{1-s}_1\right)\right)
\end{align*}
This limit can be re-written as
\begin{align}
P_{err}(M)\leq\,\frac{1}{2}\,\left[\inf_{0\in[0,1]}\,Q_s\right],
 \label{Standard quantum chernoff bound}
\end{align}
where
\begin{align}
Q_s:=\,Tr \left(\rho^s_0\rho^{1-s}_1\right).
\end{align}
These expressions for the quantum Chernoff's bounds and specially the expression \eqref{Standard quantum chernoff bound}, is the general result used for any scheme of quantum illumination. In particular, it used to evaluate the probability of errors in Lloyd's quantum illumination \cite{Lloyd2008} and for Gaussian quantum illumination \cite{Pirandola Lloyd,Tan}.

In Gaussian quantum illumination, a simplified form the Chernoff's bound consist to the evaluation of the case $s=1/2$ (Bhattacharyaya's bound),
\begin{align}
P_B(M):=\frac{1}{2}\,\left[Tr(\rho^{1/2}_0\,\rho^{1/2}_1)\right]^M.
\label{Bahttacharyya}
\end{align}
This expression is formally easier to determine, but it leads to a bound $P_{err}\leq P_B(M)$ which is weaker than Chernoff's bound \eqref{Standard quantum chernoff bound}. However, in the limit when $\rho_1$ and $\rho_0$ are very close to each other in the sense that for large $M$, $P_B$ provides an estimate for the probability of error very close to Chernoff's bound \cite{Pirandola Lloyd}. Since $P_B$ is theoretically not smaller than Chernoff's bound, to show that non-Gaussian quantum illumination presents advantage in sensitivity with respect to two mode Gaussian quantum illumination, it is enough to show that the Bahttacharyya's bound \eqref{Bahttacharyya} is smaller than Chernoff's bound for Gaussian quantum illumination in Tan et. al theory.
\subsection{Bhattacharyya's bound for three modes quantum illumination}
In the domain when $\theta \ll 1$ and $\bar{n}_2,\bar{n}_3\gg 1$ the Bhattacharyaya's bound can be computed analytically. In order to make the argument more elegant and less cumbersome we assume that $\bar{n}_2\approx \bar{n}_3$ that we denoted by $\bar{n}$.
We differentiate two scenarios. When the target is not there (Hypothesis $H_0$), the state describing the idler and the received signal is of the form
\begin{align}
\rho_0 =\,\left(U |0\rangle\langle 0| U^\dag\right)\otimes\frac{I}{\bar{n}}\otimes \frac{I}{\bar{n}},
\label{quantum state H0}
\end{align}
where the first two factors are attached to the signal photons, while the third factor describes the idler photon.

In the case when target is there, the state at the receiver is a random mix between idler-signal and idler-background states, the mix determined by the reflectivity coefficient $\eta$. If the signal is not affected by the scattering with the target, then the state is of the form
\begin{align}
 \rho_1 =\,(1-\eta)\rho_0\,+\eta \rho_\psi.
\label{quantum state H1}
\end{align}
This state can be expressed in a compact way,
\begin{align*}
\rho_1=\,\left( U |0\rangle\langle 0|U^\dag\right)\otimes (1-\eta)\frac{I}{\bar{n}}\otimes \frac{I}{\bar{n}}+\,\eta \,{\bf U}|000\rangle\langle 000 |{\bf U}^\dag .
\end{align*}

For the estimation of upper bounds of the probability of error, in particular, the Bhattacharyya's bound, it is enough to calculate the trace $[Tr(\rho^{1/2}_0\,\rho^{1/2}_1)]$. For Gaussian quantum illumination, well established calculational kit-tool have been developed \cite{Pirandola Lloyd}. But in the case on hand in the present paper, that are non-Gaussian states, the Gaussian theory as developed in \cite{Pirandola Lloyd} is not applicable. Therefore, we have proceed to evaluate the trace from first principles. The details are discussed in {\it Appendix} \ref{Evaluation of the Trace}. Disregarding terms proportional to $1/\bar{n}^2$ with respect to terms proportional to $\eta$, but still adopting the condition $\eta\ll 1$, the trace $Tr \big[{\rho}^{1/2}_0\,\rho^{1/2}_1\big]$ is of the form
\begin{align*}
Tr \big[{\rho}^{1/2}_0\,\rho^{1/2}_1\big]  =\,1-\frac{\eta^{1/2}}{\bar{n}}+\mathcal{O}(\theta^2\eta^{1/2}).
\end{align*}
The probability of error for $M$ independent identical experiments using three photons non-Gaussian states takes the form
\begin{align}
P^{3\gamma}_{err}(M)\leq \frac{1}{2}\,\exp \left(-\frac{\,M \eta^{1/2}}{\bar{n}}\right),
\label{probability of error for Non-Gaussian three modes}
\end{align}
that besides the above conditions, is approximately valid disregarding terms of order $\frac{1}{\bar{n}^2}$ or higher, in the regime where $\quad \bar{n}\gg 1$, $\eta \ll 1, \,\theta\ll 1$ and $1/\bar{n}^2\ll \,\eta$.
Remarkably, the probability of error is independent of the average number per mode $n_s=\,\theta^2$.

We aim to compare the bound \eqref{probability of error for Non-Gaussian three modes} for non-Gaussian quantum illumination with three photons states with the corresponding bounds for the probability of error for Gaussian quantum illumination in a noisy background regime where $\bar{n}\gg 1$ under the conditions of \cite{Tan}. Recall that the theoretical probability of error is in this case such that
\begin{align}
P^{2\gamma}_{err} (M)  \leq  \,\frac{1}{2}\exp \left(-\frac{M\kappa\, N_S}{\bar{n}}\right).
\label{Bboundfor2m Noisy}
\end{align}
where $\kappa$ is the emitter-to-target-to-receiver transitivity that we identify with $\eta^{1/2}$. This expression is valid in the regime where $0<\kappa\ll 1$, $N_S\ll 1$ and $N_B\ll 1$.
To compare the probability of non-Gaussian quantum illumination with three modes with the probability of error of Gaussian illumination we start identifying the identify $n_s=\theta^2$ with $N_S$. This is reasonable when the signal beam of the three photon states is characterized by the average number state $n_s= \theta^2$, while in Tan et al. the state is characterized by $N_S$, the average number per mode $m$. If we assume these identifications, then we have
\begin{align*}
P^{3\gamma}_{err}(M)\leq \frac{1}{2}\,\exp \left(-\frac{ M \kappa \,}{\bar{n}}\right).
\end{align*}
Since $\theta\ll 1$, it is clear that $P^{3\gamma}_{err} (M)\ll P^{2\gamma}_{err} (M)$ as long as ${N_S}< 1$. Note that in applying \eqref{probability of error for Non-Gaussian three modes} there is the supplementary restriction $1/\bar{n}^2\ll \,\eta$. For fixed background, this is a constraint on the type of targets under consideration. Under this additional restriction,  the regime of validity of \eqref{probability of error for Non-Gaussian three modes} is compatible with the regime of validity of Tan et al. quantum illumination. Thus assuming the above identifications and under this additional restriction on the nature of the target, the probability of error exponent is increased by a factor $1/{N_S}$ with respect to Gaussian. For example, for the figures usually discussed in the literature \cite{Shapiro2019}, $N_S=0.01$, that implies an increase of a factor $100$ in the probability error exponent for non-Gaussian quantum illumination with respect to Tan et al. Gaussian quantum illumination.
\section{Conclusion} In this paper it has been shown that for quantum illumination based upon non-Gaussian three photons entangled states of the form \eqref{Approximate Non-gaussian state for 3m} there is an enhancement in sensitivity with respect to two modes Gaussian quantum illumination that is reflected in a reduction of the probability of error exponent in a factor of order $1/{N_S}$. This result is in the same vein as the theory developed in \cite{Ricardo 2021}. Given the demonstrated experimental feasibility of the three photons states with remarkably high intensities \cite{C.W. Sandbo Chang et al. 2020}, it is realistic to expect the possibility of experimental demonstrations of the advantage of three mode quantum illumination with respect to Tan et al. Gaussian quantum illumination.

\subsection*{Acknowledgements}
This work was financially supported by the Fraunhofer Institute for High Frequency Physics and Radar Techniques.

\appendix
\section{Evaluation of the trace $Tr[\rho^{1/2}_0\rho^{1/2}_1]$.}\label{Evaluation of the Trace}
The trace $Tr [\rho^{1/2}_0\,\rho^{1/2}_1]$ can be evaluated directly under certain assumptions, providing analytic, compact expressions for the probability of error detection. One key point of the calculation below is that the matrices determining $\rho_0$ and $\rho_1$ are linear combination of projector operators, except for certain operator components acting on the subspace $V=\, spam\{|000\rangle,|111\rangle\}$ that are not projectors and that are related with the entanglement of the system.

 For the square root of $\rho_0$ we find the expression
\begin{align*}
\rho^{1/2}_0=\,& \left( \cos^2\theta |0\rangle\langle 0 |+\,\sin^2\theta |1\rangle\langle 1 |-\,\imath \sin\theta \cos \theta |1\rangle \langle 0|+\,\imath \cos \theta \sin \theta |0\rangle\langle 1|\right)\otimes \\
& \otimes \frac{1}{\bar{n}}\,\left(|0\rangle\langle 0|+|1\rangle\langle 1|+\sum^{+\infty}_{k=2}| k\rangle \langle k |\right)\otimes\,\left(|0\rangle\langle 0|+|1\rangle\langle 1|+\sum^{+\infty}_{j=2}| j\rangle \langle j | \right) ,
\end{align*}
where for the first factor we have used that
$ \left(U |0\rangle\langle 0| U^\dag\right)^{1/2} =\,\pm\, U |0\rangle\langle 0| U^\dag $ and taking the positive sign.
In the subsequent calculations we will choose the positive sign in the square root.

The evaluation of $\rho^{1/2}_1$ is more involved. We first re-cast $\rho_1$ in a convenient way,
\begin{align*}
\rho_1 =\,&\cos^2\theta \left(\frac{(1-\eta)}{\bar{n}^2}+\eta\right)|000\rangle\langle 000|+\,\sin^2\theta \left(\frac{(1-\eta)}{\bar{n}^2}+\eta\right)|111\rangle\langle 111| -\\
& -\imath\,\eta\,\cos\theta\sin\theta |111\rangle \langle 000| +\imath\,\eta\sin\theta\cos\theta \,|000\rangle\langle 111 |+\\
& + \frac{(1-\eta)}{\bar{n}^2}\left(\sin^2\theta(|100\rangle\langle 100|+|101\rangle\langle 101|+ |110\rangle\langle 110|)\,+\cos^2\theta(|011\rangle\langle 011| +\,|001\rangle\langle 001|
+|010\rangle\langle 010|)\right)\\
& + \frac{(1-\eta)}{\bar{n}^2}\left(\left(U |0\rangle\langle 0| U^\dag\right)\otimes(|0\rangle\langle 0|+|1\rangle\langle1|)\otimes\left(\sum^{+\infty}_{j=2}| j\rangle \langle j |\right) \right)+\\
& + \frac{(1-\eta)}{\bar{n}^2}\left(\left(U |0\rangle\langle 0| U^\dag\right)\otimes\left(\sum^{+\infty}_{j=2}| j\rangle \langle j |\right)\otimes (|0\rangle\langle 0|+|1\rangle\langle1|) \right)+\\
& + \frac{(1-\eta)}{\bar{n}^2}\left(\left(U |0\rangle\langle 0| U^\dag\right)\otimes\left(\sum^{+\infty}_{j=2}| j\rangle \langle j |\right)\otimes \left(\sum^{+\infty}_{k=2}| k\rangle \langle k |\right) \right) .
\end{align*}
We make now the approximation of disregarding terms proportional to $1/\bar{n}^2$ with respect terms proportional to $\eta$. Although this approximation is not essential and constrains the scope of our calculation, it is useful to provide clean analytical expressions. Also, we consider the condition $\eta\ll 1$ and disregard $\eta$ with respect to $1$ when they appear in an addition. Under these approximation, we have
\begin{align*}
\rho_1 \approx\,\,&\eta\left(\cos^2\theta |000\rangle\langle 000|+\,\sin^2\theta |111\rangle\langle 111| -\,\imath\,\cos\theta\sin\theta |111\rangle \langle 000| +\imath\sin\theta\cos\theta \,|000\rangle\langle 111 |\right)+\\
& + \frac{1}{\bar{n}^2}\left( \left(U |0\rangle\langle 0| U^\dag\right)\otimes(|0\rangle\langle 0|+|1\rangle\langle1|)\otimes\left(\sum^{+\infty}_{j=2}| j\rangle \langle j |\right) \right)+\\
& + \frac{1}{\bar{n}^2}\left(\left(U |0\rangle\langle 0| U^\dag\right)\otimes\left(\sum^{+\infty}_{j=2}| j\rangle \langle j |\right)\otimes (|0\rangle\langle 0|+|1\rangle\langle1|) \right)+\\
& + \frac{1}{\bar{n}^2}\left(\left(U |0\rangle\langle 0| U^\dag\right)\otimes\left(\sum^{+\infty}_{j=2}| j\rangle \langle j |\right)\otimes \left(\sum^{+\infty}_{k=2}| k\rangle \langle k |\right)\right) .
\end{align*}
The first line can be re-cast in a shorter way, since it is equal to $\eta \,{\bf U}|000\rangle\langle 000 |{\bf U}^\dag $,
\begin{align*}
\rho_1 \approx\,\,&\eta{\bf U}|000\rangle\langle 000 |{\bf U}^\dag\\
& + \frac{1}{\bar{n}^2}\left( \left(U |0\rangle\langle 0| U^\dag\right)\otimes(|0\rangle\langle 0|+|1\rangle\langle1|)\otimes\left(\sum^{+\infty}_{j=2}| j\rangle \langle j |\right) \right)+\\
& + \frac{1}{\bar{n}^2}\left(\left(U |0\rangle\langle 0| U^\dag\right)\otimes\left(\sum^{+\infty}_{j=2}| j\rangle \langle j |\right)\otimes (|0\rangle\langle 0|+|1\rangle\langle1|) \right)+\\
& + \frac{1}{\bar{n}^2}\left(\left(U |0\rangle\langle 0| U^\dag\right)\otimes\left(\sum^{+\infty}_{j=2}| j\rangle \langle j |\right)\otimes \left(\sum^{+\infty}_{k=2}| k\rangle \langle k |\right)\right) .
\end{align*}
Since most of the components are in terms of orthogonal projectors to the subspace $V=\{|000\rangle,|111\rangle\}$, the square root $\rho^{1/2}$ can be evaluated easily. Considering that each component brings its own possible two signs for each of the square roots. One needs to choose the signs consistent with the possible interpretations of the trace $Tr[\rho^{1/2}_0\,\rho^{1/2}_1]<1$ for the evaluation of the Chernoff's bounds. Since we have the square roots, he square root
\begin{align*}
({\bf U}|000\rangle\langle 000 |{\bf U}^\dag )^{1/2}=\pm {\bf U}|000\rangle\langle 000 |{\bf U}^\dag
\end{align*}
according to the above criterion, we will choose the minus sign, while for the rest of components we choose the same positive sign than the analogous components in $\rho_0$.  Then we have
\begin{align*}
\rho^{1/2}_1 \approx\,&-\eta^{1/2}\,\left(\cos^2\theta |000\rangle\langle 000|+\,\sin^2\theta |111\rangle\langle 111| -\,\imath\,\cos\theta\sin\theta |111\rangle \langle 000| +\imath\sin\theta\cos\theta \,|000\rangle\langle 111 |\right)+\\
& + \frac{1}{\bar{n}^2}\left( \left(U |0\rangle\langle 0| U^\dag\right)\otimes(|0\rangle\langle 0|+|1\rangle\langle1|)\otimes\left(\sum^{+\infty}_{j=2}| j\rangle \langle j |\right) \right)+\\
& + \frac{1}{\bar{n}^2}\left(\left(U |0\rangle\langle 0| U^\dag\right)\otimes\left(\sum^{+\infty}_{j=2}| j\rangle \langle j |\right)\otimes (|0\rangle\langle 0|+|1\rangle\langle1|) \right)+\\
& + \frac{1}{\bar{n}^2}\left(\left(U |0\rangle\langle 0| U^\dag\right)\otimes\left(\sum^{+\infty}_{j=2}| j\rangle \langle j |\right)\otimes \left(\sum^{+\infty}_{k=2}| k\rangle \langle k |\right)\right) .
\end{align*}
 The calculation follows now by multiplying $\rho^{1/2}_0$ and $\rho^{1/2}_1$ and taking the trace. Since we have choose the same sign in the square roots of the projectors orthogonal to $V$, we have
 \begin{align*}
 \rho^{1/2}_1=\,\rho^{1/2}_0-\,\eta^{1/2}{\bf U}|000\rangle\langle 000 |{\bf U}^\dag -\frac{1}{\bar{n}}\,\left (|0\rangle\langle0|+|1\rangle\langle 1|)\otimes (|0\rangle\langle 0|+|1\rangle\langle1|)\otimes\left(U |0\rangle\langle 0| U^\dag\right) \right) .
 \end{align*}
The last terms, being finite and proportional to the factor $1/\bar{n}$, when multiplied by $\rho$ bring a factor of order $1/\bar{n}^2$, that it is disregarded.
Therefore, we have
\begin{align*}
Tr[\rho^{1/2}_0\,\rho^{1/2}_1]=\,Tr[\rho^{1/2}_0\,(\rho^{1/2}_0-\,\eta^{1/2}{\bf U}|000\rangle\langle 000 |{\bf U}^\dag )]=\,1-\,Tr[\rho^{1/2}_0\,\eta^{1/2}{\bf U}|000\rangle\langle 000 |{\bf U}^\dag ].
\end{align*}
The task now is to evaluate the part of the product where the subspace $\{|000\rangle, |111\rangle\}$. In particular, we need to evaluate the part of the trace
\begin{align*}
& Tr[\rho^{1/2}_0\,\eta^{1/2}{\bf U}|000\rangle\langle 000 |{\bf U}^\dag ]=\, Tr[\frac{1}{\bar{n}}\,\eta^{1/2}\,\left( \cos^2\theta |0\rangle\langle 0 |+\,\sin^2\theta |1\rangle\langle 1 |-\,\imath \sin\theta \cos \theta |1\rangle \langle 0|+\,\imath \cos \theta \sin \theta |1\rangle\langle 0|\right)\otimes\\
&\otimes(|0\rangle\langle 0|+|1\rangle\langle 1|)\otimes(|0\rangle\langle 0|+|1\rangle\langle 1|\otimes\left( \cos^2\theta |0\rangle\langle 0 |+\,\sin^2\theta |1\rangle\langle 1 |-\,\imath \sin\theta \cos \theta |1\rangle \langle 0|+\,\imath \cos \theta \sin \theta |1\rangle\langle 0|\right)\cdot\\
&\cdot \left(\cos^2\theta |000\rangle\langle 000|+\,\sin^2\theta |111\rangle\langle 111| -\,\imath\,\cos\theta\sin\theta |111\rangle \langle 000| +\imath\sin\theta\cos\theta \,|000\rangle\langle 111 |\right)]\\
& = \,\frac{\eta^{1/2}}{\bar{n}}+\mathcal{O}(\theta^3).
\end{align*}
Disregarding terms proportional to $1/\bar{n}^2$ with respect to $\eta$, but still adopting the condition $\eta\ll 1$, is
\begin{align}
Tr \big[{\rho}^{1/2}_0\,\rho^{1/2}_1\big]  =\,1-\frac{\eta^{1/2}}{\bar{n}}+\mathcal{O}(\theta^2\eta^{1/2}),
\label{expression for the trace}
\end{align}
where we have made us of the approximation $\sin (\theta)\, \approx \theta$, since $\theta$ is small and we have disregarded terms of  order $\mathcal{O}(\theta^3)$ and of order $1/\bar{n}^2$ compared with terms of order $\eta$.

\end{document}